\documentclass[12pt]{article}
\usepackage{epsf}
\usepackage{psfig}
\usepackage{times}
\usepackage{latexsym}
\usepackage{amsbsy}
\usepackage{ifthen}

\usepackage[dvips]{epsfig}

\begin{document}

\def\##1{\bf{ #1}}
\def\=#1{\underline{\underline #1}}

\def\eps{\epsilon}
\def\epso{\epsilon_0}
\def\muo{\mu_0}
\def\ko{k_0}
\def\kosq{k_0^2}
\def\lambdao{\lambda_0}
\def\etao{\eta_0}
\def\.{\mbox{ \tiny{$^\bullet$} }}

\def\curl{\nabla\times}
\def\div{\nabla \mbox{ \tiny{$^\bullet$} }}

\def\ux{\#{u}_x}
\def\uy{\#{u}_y}
\def\uz{\#{u}_z}
\def\up{\#{u}_+}
\def\um{\#{u}_-}

\def\le{\left(}
\def\ri{\right)}
\def\les{\left[}
\def\ris{\right]}
\def\lec{\left\{}
\def\ric{\right\}}

\def\c#1{\cite{#1}}
\def\l#1{\label{#1}}
\def\r#1{(\ref{#1})}

\def\Bdc{{\bf B}_{dc}}
\def\Nt{\tilde{N}}

\newcommand{\mat}[1] {\left[\begin{array}{cccc}
            #1 \end{array}\right]}

\newcommand{\mb}[1]{\mbox{\boldmath$\bf#1$}}

{\bf On Onsager Relations and Linear Electromagnetic Materials}
\\
\bigskip

A. Lakhtakia,
 CATMAS---Computational \& Theoretical
Materials Science Group, Department of Engineering Science \&
Mechanics, Pennsylvania State University, University Park, PA
16802--6812, USA.\\ E--mail: akhlesh@psu.edu\\

\bigskip

R.A. Depine, Grupo de Electromagnetismo Aplicado\\
Departamento de 
F\'{\i}sica, Universidad de 
Buenos Aires \\
Ciudad Universitaria, Pabell\'{o}n I,
1428 Buenos Aires, Argentina\\
E--mail: rdep@df.uba.ar\\

{\bf We investigated the Onsager relations in the context of electromagnetic
constitutive relations of  linear, homogeneous 
materials. We determined that application of the Onsager relations to the constitutive equations relating $\#P$ and $\#M$ to both $\#E$ and $\#B$
is in accord with Lorentz reciprocity as well as the Post constraint.
}

\section{Introduction}
In two seminal papers published in 1931 \c{Ons1,Ons2}, with the assumption
of microscopic reversibility, Onsager derived a set of reciprocity relations applicable to coupled linear
phenomenons at macroscopic length scales. Fourteen years later, Casimir \c{Cas} improved the
foundations of the Onsager relations. Initially considered applicable to purely instantaneous phenomenons~---~or, at least,
when ``time--lag can be neglected'' \cite[p. 419]{Ons1}~---~the  Onsager relations widened in scope as a result of  the
fluctuation--dissipation theorem \c{CG52} to time--harmonic phenomenons \c{Call52}.
Sections 123--125 of the famous
textbook of Landau and Lifshitz on statistical physics provide a lucid introduction to the Onsager
relations \c{LL}, but we also recommend a perusal of a classic monograph by de Groot \c{deG}. A modern appraisal
has been provided by Berdichevsky \c{Ber}, whose paper motivated the work leading to this communication.

Our focus is  the {\em correct\/} application of the Onsager relations for linear electromagnetic materials. This
issue can be traced back to a 1973 paper by Rado \c{Rado}. This paper contains a major conflict between a consequence of the assumption
of material response without any delay whatsoever and the Onsager relations as expounded by Callen {\em et al.\/} \c{Call52}. The former
is definitely a noncausal assumption in electromagnetism \c{WL96, Tip}, leading to false symmetries between the electromagnetic
constitutive parameters \c{L94}. Furthermore, Rado considered $\#E$ and $\#H$ as primitive fields, but 
$\#E$ and $\#B$ are taken to be the primitive fields in {\em modern electromagnetism\/} \c{DZ99,Jackson,Post03}.
To the best of our knowledge, no other {\em original\/} investigation of the Onsager relations in electromagnetism
exists. 

Due to  the currently increasing emphasis on  engineered nanomaterials \c{NRC, Nalwa} and complex electromagnetic materials \c{SL00,WL03},
it is imperative that the application of fundamental principles (such as the Onsager relations) be  carefully examined
with modern terminology. Accordingly, in the following sections, we first review the Onsager relations in general. Then we apply the 
Onsager relations to the electromagnetic
constitutive relations of  linear, homogeneous, bianisotropic
materials. We show that a na\"ive application to constitutive equations relating $\#D$ and $\#H$ to both $\#E$ and $\#B$
yields unphysical results, but that application to constitutive equations relating $\#P$ and $\#M$ to both $\#E$ and $\#B$
is in accord with Lorentz reciprocity \c{Kong} as well as the Post constraint \c{WL98,L04}.

\section{Onsager relations}
Let us consider the linear macroscopic constitutive equations
\begin{equation}
\label{eqcon1}
L_m=\sum_{n=1}^N\,\Phi_{mn}\,F_n\,,\quad m\in[1,\,N]\,,
\end{equation}
where $N>1$, $L_m$ are the {\em Onsager fluxes} and $F_m$ are the {\em Onsager
forces\/}. The Onsager relations deal with the constitutive parameters $\Phi_{mn}$.

The derivation of the Onsager relations proceeds with the postulation of $N$ state variables $a_n$, $n\in[1,N]$. 
The state variables are divided into two groups.
The first $\Nt\leq N$ state variables are supposed to be {\em even\/} and the remaining $N-\Nt$ 
state variables are supposed to be {\em odd\/} with
respect to a reversal
of velocities of the microscopic particles constituting the linear medium; in other words,
\begin{eqnarray}
&& \nonumber
\overline{a_m(t)a_n(t+\tau)}=\overline{a_m(t)a_n(t-\tau)}\,,
\\
&&\quad
{\rm if}\left\{
\begin{array} {c}
m\in[1,\,\Nt] \, {\rm and}\, n\in[1,\,\Nt] \\
{\rm or}\\
m\in[\Nt+1,\,N] \, {\rm and}\, n\in[\Nt+1,\,N]
\end{array}\right.
\label{a1a}
\end{eqnarray}
and
\begin{eqnarray}
&& \nonumber
\overline{a_m(t)a_n(t+\tau)}=-\,\overline{a_m(t)a_n(t-\tau)}\,,
\\
&&\quad
{\rm if}\left\{
\begin{array} {c}
m\in[1,\,\Nt] \, {\rm and}\, n\in[\Nt+1,\,N] \\
{\rm or}\\
m\in[\Nt+1,\,N] \, {\rm and}\, n\in[1,\,\Nt]
\end{array}\right.\,,
\label{a1b}
\end{eqnarray}
where the overbar indicates averaging over time $t$ \c{Cas}.

In terms of the
state variables,
the { Onsager fluxes} are defined
as
\begin{equation}
L_m=\frac{\partial}{\partial t}\, a_m\,,\quad m \in[1,\,N]\,;
\end{equation}
the {\em Onsager forces} are defined as
\begin{equation}
F_m=-\sum_{n=1}^{N}\,g_{mn}a_n\,, \quad m \in[1,\,N]\,;
\end{equation}
and the coefficients $g_{mn}$ help define
the deviation $\Delta S$ of the entropy
from its equilibrium value as the quadratic expression \c{deG}
\begin{eqnarray}
&& \nonumber
\Delta S = -\frac{1}{2}\sum_{m=1}^{\Nt}  \sum_{n=1}^{\Nt} \, g_{mn}a_ma_n\\
&& \quad
-\frac{1}{2}\sum_{m=\Nt+1}^N\sum_{n=\Nt+1}^N\, g_{mn}a_ma_n\,.
\end{eqnarray}

In consequence of the microscopic reversibility indicated by \r{a1a}
and \r{a1b}, the constitutive parameters satisfy the Onsager
relations
\begin{eqnarray}
&& \nonumber
\Phi_{mn}=\Phi_{nm}\,,
\\
&&\quad
{\rm if}\left\{
\begin{array} {c}
m\in[1,\,\Nt] \, {\rm and}\, n\in[1,\,\Nt] \\
{\rm or}\\
m\in[\Nt+1,\,N] \, {\rm and}\, n\in[\Nt+1,\,N]
\end{array}\right.
\label{eqOn1a}
\end{eqnarray}
and
\begin{eqnarray}
&& \nonumber
\Phi_{mn}=-\Phi_{nm}\,,
\\
&&\quad
{\rm if}\left\{
\begin{array} {c}
m\in[1,\,\Nt] \, {\rm and}\, n\in[\Nt+1,\,N] \\
{\rm or}\\
m\in[\Nt+1,\,N] \, {\rm and}\, n\in[1,\,\Nt]
\end{array}\right.\,.
\label{eqOn1b}
\end{eqnarray}
In an external magnetostatic field $\Bdc$, \r{eqOn1a} and \r{eqOn1b} are modified to
\begin{eqnarray}
&& \nonumber
\Phi_{mn}(\Bdc)=\Phi_{nm}(-\Bdc)\,,
\\
&&\quad
{\rm if}\left\{
\begin{array} {c}
m\in[1,\,\Nt] \, {\rm and}\, n\in[1,\,\Nt] \\
{\rm or}\\
m\in[\Nt+1,\,N] \, {\rm and}\, n\in[\Nt+1,\,N]
\end{array}\right.
\label{eqOn2a}
\end{eqnarray}
and
\begin{eqnarray}
&& \nonumber
\Phi_{mn}(\Bdc)=-\Phi_{nm}(-\Bdc)\,,
\\
&&\quad
{\rm if}\left\{
\begin{array} {c}
m\in[1,\,\Nt] \, {\rm and}\, n\in[\Nt+1,\,N] \\
{\rm or}\\
m\in[\Nt+1,\,N] \, {\rm and}\, n\in[1,\,\Nt]
\end{array}\right.\,,
\label{eqOn2b}
\end{eqnarray}
respectively.

\section{Application to Linear Electromagnetism}

\subsection{Constitutive Equations for $\#D$ and $\#H$}
Let us now consider a linear, homogeneous, bianisotropic medium. Its
constitutive equations can be written in a cartesian coordinate system as
\begin{equation}
\label{eq1}
\left.\begin{array}{ll}
D_j  = \sum_{k=1}^3\,\epsilon_{jk}\circ E_k  +\xi_{jk}\circ B_k \\[5pt]
H_j  = \sum_{k=1}^3\,\zeta_{jk}\circ E_k  +\nu_{jk}\circ B_k
\end{array}\right\}\,, \quad j\in[1,3]\,.
\end{equation} 
We have adopted here the
modern view of electromagnetism wherein $\#E$ and $\#B$ are
the primitive fields while $\#D$ and $\#H$ are the induction fields \c{DZ99,Jackson,Post03}.
The operation $\circ$ indicates a temporal convolution operation in the time domain,
and simple multiplication in the frequency domain \c{Weigl03}.

Now, $\#D$ and $\#E$ are even, but $\#H$ and $\#B$ are odd, with respect to time--reversal. With that in mind,
we can rewrite \r{eq1} compactly as
\begin{equation}
\label{eqbon2}
Q_m=\sum_{n=1}^N\,\Lambda_{mn}\circ F_n\,,\quad m\in[1,\,N]\,,
\end{equation}
where 
$F_m=E_m$,
$F_{m+3}=B_m$, $Q_m=D_m$ and
$Q_{m+3}=H_m$ for $m\in\left[1,\,3\right]$; furthermore, $\Nt=3$ and $N=6$.

With the assumption of microscopic reversibility, 
application of  the Onsager relations \r{eqOn2a} and \r{eqOn2b} yields
the following symmetries:
\begin{equation}
\label{eq7a}
\left.\begin{array}{ll}
\Lambda_{mn}(\Bdc)=\Lambda_{nm}(-\Bdc)\,,&\,\, m\in\left[1,\,3\right]\,,\,n\in\left[1,\,3\right]\\[4pt]
\Lambda_{mn}(\Bdc)=\Lambda_{nm}(-\Bdc)\,,&\,\, m\in\left[4,\,6\right]\,,\,n\in\left[4,\,6\right]\\[4pt]
\Lambda_{mn}(\Bdc)=-\Lambda_{nm}(-\Bdc)\,,&\,\, m\in\left[1,\,3\right]\,,\,n\in\left[4,\,6\right]\\[4pt]
\end{array}\right\}\,.
\end{equation}
Equations \r{eq7a} imply that
\begin{equation}
\label{eq7b}
\left.\begin{array}{ll}
\epsilon_{jk}(\Bdc) = \epsilon_{kj}(-\Bdc)\\[4pt]
\nu_{jk}(\Bdc) = \nu_{kj}(-\Bdc)\\[4pt]
\xi_{jk}(\Bdc) = -\zeta_{kj}(-\Bdc)
\end{array}\right\}\,.
\end{equation}

\subsection{Constitutive Equations for $\#P$ and $\#M$}
When considering a material medium, as distinct from matter--free space (i.e., vacuum), the presence of matter is
indicated by the
the polarization ${\#P}={\#D}-\epsilon_o{\#E}$ and the magnetization ${\#M}=\mu_o^{-1}{\#B}-{\#H}$,
where $\epsilon_o$ and $\mu_o$ are the permittivity and the permeability of
matter--free space. Linear constitutive equations for $\#P$ and $\#M$ can be stated as
\begin{equation}
\label{eq3}
\left.\begin{array}{ll}
P_j  = \sum_{k=1}^3\,\chi^{(1)}_{jk}\circ E_k  +\chi^{(2)}_{jk}\circ B_k \\[5pt]
M_j  =\sum_{k=1}^3\, \chi^{(3)}_{jk}\circ E_k  +\chi^{(4)}_{jk}\circ B_k 
\end{array}\right\}\,, \quad j\in[1,3]\,,
\end{equation} 
where
\begin{equation}
\label{eq4}
\left.\begin{array}{ll}
\epsilon_{jk} =\epsilon_o\delta_{jk}+\chi^{(1)}_{jk}\\[5pt]
\nu_{jk}=\mu_o^{-1}\delta_{jk}-\chi^{(4)}_{jk}\\[5pt]
\xi_{jk}=\chi^{(2)}_{jk}\\[5pt]
\zeta_{jk}=-\chi^{(3)}_{jk}
\end{array}\right\}\,,
\end{equation} 
and $\delta_{jk}$ is the Kronecker delta function.

As $\#P$ is even but $\#M$ is  odd with respect to time--reversal,
we can rewrite \r{eq3} as
\begin{equation}
\label{eqcon2}
R_m=\sum_{n=1}^N\,\Psi_{mn}\circ F_n\,,\quad m\in[1,\,N]\,,
\end{equation}
where 
$R_m=P_m$ and
$R_{m+3}=M_m$ for $m\in\left[1,\,3\right]$. 
As the
microscopic processes underlying the constitutive parameters in \r{eqcon2} are reversible,
$\Psi_{mn}$ must satisfy \r{eqOn2a} and \r{eqOn2b}; thus,
\begin{equation}
\label{eq9a}
\left.\begin{array}{ll}
\Psi_{mn}(\Bdc)=\Psi_{nm}(-\Bdc)\,,&\,\, m\in\left[1,\,3\right]\,,\,n\in\left[1,\,3\right]\\[4pt]
\Psi_{mn}(\Bdc)=\Psi_{nm}(-\Bdc)\,,&\,\, m\in\left[4,\,6\right]\,,\,n\in\left[4,\,6\right]\\[4pt]
\Psi_{mn}(\Bdc)=-\Psi_{nm}(-\Bdc)\,,&\,\, m\in\left[1,\,3\right]\,,\,n\in\left[4,\,6\right]\\[4pt]
\end{array}\right\}\,,
\end{equation}
whence the symmetries
\begin{equation}
\label{eq5b}
\left.\begin{array}{ll}
\chi^{(1)}_{jk}(\Bdc) = \chi^{(1)}_{kj}(-\Bdc)\\[5pt]
\chi^{(4)}_{jk}(\Bdc) = \chi^{(4)}_{kj}(-\Bdc)\\[5pt]
\chi^{(2)}_{jk}(\Bdc) = -\chi^{(3)}_{kj}(-\Bdc)
\end{array}\right\}\,
\end{equation}
are predicted by the Onsager relations as the macroscopic consequences of microscopic reversibility.

\subsection{The Conflict}

Equations \r{eq5b} imply that
\begin{equation}
\label{eq5c}
\left.\begin{array}{ll}
\epsilon_{jk}(\Bdc) = \epsilon_{kj}(-\Bdc)\\[4pt]
\nu_{jk}(\Bdc) = \nu_{kj}(-\Bdc)\\[4pt]
\xi_{jk}(\Bdc) = \zeta_{kj}(-\Bdc)
\end{array}\right\}\,,
\end{equation}
by virtue  of \r{eq4}. 

But \r{eq5c}$_3$ disagrees completely with \r{eq7b}$_3$. Let us reiterate that
both \r{eq7b}$_3$ and \r{eq5c}$_3$
come about from the application of the Onsager relations, contingent upon the assumption
of microscopic reversibility. Yet, at most, only one of the two must be correct.

\subsection{Resolution of the Conflict}

Onsager's own papers help resolve the conflict. His papers were concerned with motion of microscopic
particles, and he considered his work to hold true for heat conduction, gaseous diffusion and related transport problems.
The Onsager forces must be causative agents, while the Onsager fluxes must be directly concerned with particulate
motion. This understanding is reinforced by subsequent commentaries \c{LL,deG}.

Therefore, in order to {\em correctly} exploit the Onsager relations
in electromagnetics, we must isolate those parts of $\#D$ and $\#H$ which
indicate the presence of a {\em material}, because microscopic processes cannot occur in matter--free space (i.e., vacuum). 
The matter--indicating parts of
$\#D$ and $\#H$ are $\#P$ and $\#M$. {\em Hence, \r{eq5c} must be accepted and  \r{eq7b} must be discarded.}

With $\Bdc=\#0$, the symmetries \r{eq5c} coincide~---~unlike  \r{eq7b}~---~with those mandated
by Lorentz reciprocity \cite[Eqs. 23]{Kong}. Also unlike  \r{eq7b}, the symmetries \r{eq5c}
are compatible with the Post constraint \c{WL98,L04}
\begin{equation}
\sum_{j=1}^3\,\xi_{jj}=\sum_{j=1}^3\,\zeta_{jj}\,
\end{equation}
which must be satisfied by all (i.e., Lorentz--reciprocal as well as Lorentz--nonreciprocal) linear materials. These
two well--known facts also support our decision to discard \r{eq7b} in favor of  \r{eq5c}.

\section{Concluding Remarks}
In this communication, we first reviewed the Onsager relations which delineate the macroscopic consequences
of microscopic reversibility in linear materials. Then we applied the relations to the electromagnetic
constitutive relations of  homogeneous bianisotropic
materials. We determined that a na\"ive application to constitutive equations relating $\#D$ and $\#H$ to both $\#E$ and $\#B$
yields unphysical results, but that application to constitutive equations relating $\#P$ and $\#M$ to both $\#E$ and $\#B$
is in accord with Lorentz reciprocity as well as the Post constraint.

\end{document}